\documentclass{article}
\usepackage{spconf,amsmath,graphicx, cite, blindtext}
\usepackage[dvipsnames]{xcolor}
\usepackage{booktabs}
\usepackage{xcolor,colortbl}
\usepackage{diagbox}

\usepackage{hyperref}
\usepackage[all]{hypcap}
\hypersetup{
    colorlinks = false,
    linkbordercolor = {white}
}

\title{SPIKING NEURAL NETWORKS TRAINED WITH BACKPROPAGATION FOR LOW POWER NEUROMORPHIC IMPLEMENTATION OF VOICE ACTIVITY DETECTION}

\name{Flavio Martinelli\textsuperscript{$*$},
      Giorgia Dellaferrera\textsuperscript{$*$},
      Pablo Mainar\textsuperscript{$\dagger$},
      Milos Cernak\textsuperscript{$\dagger$}}
\address{\textsuperscript{$*$}\'Ecole Polytechnique F\'ed\'erale de Lausanne (EPFL), Switzerland\\
         \textsuperscript{$\dagger$}Logitech Europe S.A., Lausanne, Switzerland}

\begin{document}

\maketitle

\begin{abstract}

Recent advances in Voice Activity Detection (VAD) are driven by artificial and Recurrent Neural Networks (RNNs), however, \mbox{using} a VAD system in battery-operated devices requires further power efficiency. This can be achieved by \mbox{neuromorphic} hardware, which enables Spiking Neural Networks (SNNs) to perform inference at very low energy consumption. 
Spiking networks are characterized by their ability to process information efficiently, in a sparse cascade of binary events in time called spikes. However, a big performance gap separates artificial from spiking networks, mostly due to a lack of powerful SNN training algorithms. To overcome this problem we exploit an SNN model that can be recast into a recurrent network and trained with known deep learning techniques. We describe a training procedure that achieves low spiking activity and apply pruning algorithms to remove up to 85\% of the network connections with no performance loss. The model competes with state-of-the-art performance at a fraction of the power consumption comparing to other methods.

\end{abstract}


\section{Introduction}
\label{sec:intro}
Voice Activity Detection (VAD) is the preliminary, gating, step in most of speech processing applications. It tackles the non-trivial problem of discriminating speech signals from environmental background noise, the latter being the main source of performance impairment of \mbox{Automatic} Speech Recognition (ASR) systems \cite{ramirez2007voice}.
The issue of power consumption is particularly relevant if such systems are \mbox{embedded} in battery-powered devices, therefore an embedded always-on voice detector should be evaluated considering both detection performance and energy consumption; such criteria are in contrast between each other.

Contemporary VAD approaches range from the digital signal processing ones \cite{li2004complexity, benyassine1997itu, sohn1999statistical, ramirez2004efficient} to data-driven (trained) ones,  such as Gaussian Mixture Models (GMM) \cite{ng2012developing, ghaemmaghami2015complete}, and recent deep learning approaches using Artificial, Convolutional \cite{silva2017exploring} or Recurrent Neural Networks \cite{gelly2017optimization} (ANNs, CNNs or RNNs). Notably, ANN-based techniques outperform all previous methods but these models are computationally heavy limiting their deployment on battery-powered devices. This problem is often targeted by reducing the size and complexity of the ANN models with various compressing techniques \cite{cheng2017survey} and \mbox{dedicated} low power hardware.

In this work, we propose an energy-efficient VAD algorithm based on spiking networks that respect the implementability constraints of the current state-of-the-art neuromorphic chips. 
Neuromorphic hardware's main strength is the very low power consumption (c.f. Table~1 of \cite{frenkel20180}), making it suitable for embedded machine learning tasks.
Spiking neurons, in comparison to ANN units, represent more closely biological neurons, are more complex and show higher computational capacity \cite{maass1997networks}. However,
spiking networks still lack a powerful training algorithm
such as what backpropagation is for ANNs. Moreover, at the moment,
\mbox{biologically} derived learning rules for SNNs do not match performances of ANNs or other SNN training
techniques \cite{tavanaei2018deep}.

Our model is inspired by the recently proposed formulation of discrete-time SNNs as Recurrent Neural Networks \cite{neftci2019surrogate, bellec2018long, essera2016convolutional}, overcoming the non-differentiability of the spike activation function by using a surrogate gradient that allows the backpropagation of errors. RNN is to be intended as the large class of networks sharing time dynamics and recurrent connections, as stated also in \cite{neftci2019surrogate}.
This formulation of SNNs into RNNs is a key component in our work since it broadens the pool of available optimization techniques to the ones developed for deep learning. 

With respect to ANN to SNN conversion methods \cite{rueckauer2017conversion}, our models exploit the neurons' time dynamics, explicitly adding another dimension to the encoding space of information. Such information is encoded in the \mbox{synchronicity} of the spikes, thus reducing the number of binary events needed to represent the signal; hence reducing energy consumption.

To deploy such models into neuromorphic hardware one must take into consideration some architectural constraints, mainly the limited number of connections per \mbox{neuron}. We assess this issue by showing that 'lottery tickets' \mbox{subnetworks} \cite{frankle2018the} can be found in our models, reducing the number of connections to 15\% the original without decrease of performance.

\section{Methods}

\subsection{Spiking network model}
\label{sec:model}
The key insight of the spiking network proposed by Neftci \textit{et al.} \cite{neftci2019surrogate} and Bellec \textit{et al.} \cite{bellec2018long} is the reformulation of the Leaky Integrate and Fire (LIF) model with synaptic currents \cite{gerstner2014neuronal} into discrete-time equations and its equivalence to a Recurrent Neural Network with binary activation functions and reset functions. Here we briefly summarize the link between continuous-time LIF and the model proposed in \cite{neftci2019surrogate}.

The LIF model is a widely used neuron model implemented in most of neuromorphic chips, it describes the state of neuron $i$ with voltage membrane $V_i(t)$ and synaptic current $I_i(t)$ variables, evolving through time according to these differential equations:
\begin{align}
\label{eqn:LIF}
\begin{split}
\tau_{mem} \dot{V_i}(t) &= -V_i(t) + RI_i(t) , 
\\
\tau_{syn} \dot{I_i}(t) &= -I_i(t) + \sum_jw_{ij}s_j(t)
\end{split}
\end{align}
where $s_j(t) = \sum_k \delta(t-t_k)$ are input spikes coming from the presynaptic neuron $j$; $\delta$ being the Dirac's delta distribution. 

The synaptic current variable $I_i(t)$ is incremented each time a spike reaches neuron $i$ from neuron $j$ (proportionally to the synaptic weight $w_{ij}$)  and follows the behaviour of a leaky integrator with decay constant $\tau_{syn}$. The voltage variable also follows the same behaviour with constant $\tau_{mem}$. A neuron emits a spike whenever its membrane voltage potential reaches a set threshold $\theta$, then, immediately after the spike, the neuron's potential is reset to zero. Eqs.\ref{eqn:LIF} can be approximated in discrete timesteps with $R=\theta=1$ and $\Delta T$ sufficiently small:
\begin{align}
\label{eqn:discreteLIF}
\begin{split}
V_i(t+\Delta T) &= \alpha V_i(t) + I_i(t) - S_i(t) , 
\\
I_i(t+\Delta T) &= \beta I_i(t) + \sum_jw_{ij}S_j(t)
\end{split}
\end{align}
with decay constants $\alpha$ and $\beta$ equivalent to $exp(-\frac{\Delta T}{\tau_{mem}})$ and $exp(-\frac{\Delta T}{\tau_{syn}})$ respectively. Spikes can be now expressed as result of a binary activation function applied to the membrane potential of the neuron: $S_i(t) = \Theta(V_i(t)-\theta)$; $\Theta$ being the Heaviside function. Finally, the reset mechanism is given by subtracting the spike function $S_i$ to $V_i$.

Equations \ref{eqn:discreteLIF} are mathematically equivalent to an RNN with $V,I$ as internal states and $\Theta(V-\theta)$ as activation function. This allows to train this model using the tools of deep learning, computing the gradients from eqs.\ref{eqn:discreteLIF} with Backpropagation Through Time (BPTT).

The activation function's derivative, $\dot{\Theta} = \delta$, is zero everywhere except at the origin, where it is not defined, preventing the backpropagation of errors. It is thus been proposed to substitute its derivative with the derivative of a surrogate continuous and differentiable function that allows the flow of gradient through the network \cite{shrestha2018slayer,zenke2018superspike, bellec2018long, essera2016convolutional}.
The choice for our model fell on the derivative of the fast sigmoid \cite{zenke2018superspike}:
\begin{equation}
\label{eqn:fastsig}
\frac{1}{(1 + \lambda|V_i(t) - \theta|)^2} 
\end{equation}
where $\lambda=10$ sets the steepness of the sigmoid.

The gradient is not allowed to flow through the reset of the voltage variable since it has been noticed that the reset function's large negative gradient contributes to changes in weights that force the neuron potential to be repelled from the threshold and therefore silencing the network \cite{zenkePoster}.

The model's parameters (e.g. $w$) can be trained to minimize a loss function with BPTT by using eqs.\ref{eqn:discreteLIF} in the forward pass. The backward pass is computed using the same eqs. but substituting the derivative of $S(t)$ with eq.\ref{eqn:fastsig}.

\vspace{-2pt}
\subsection{Preprocessing and temporal spike encoding}

The features given to the SNN model are extracted from audio tracks sampled at $16$ kHz. $128$ log Mel filterbank coefficients are computed on windows of $64$ ms with $75\%$ of overlap which leads to a feature vector of $128$ coefficients every $16$ ms. Each coefficient is then normalized between $0$ and $1$ with respect to the minimum and maximum value in the entire training dataset. For the testing dataset, the same procedure is performed using the training normalization coefficients and clipping every resulting value to the interval $[0,1]$.

The $128$ normalized coefficients of each audio frame, $x_{k}$, are then encoded into spike patterns of $T$ time-steps, following the rule of time-to-first-spike \cite{gerstner2014neuronal}. The higher the value the earlier the corresponding neuron $k$ will spike.
\begin{equation}\label{ttfs}
    S_k(t) = \delta[ t - \left \lfloor T\cdot(1-x_{k})  \right \rceil]
\end{equation}
where $\delta[\cdot]$ is the discrete impulse function and the quantity $ T \cdot (1-x_{k})$ is rounded to the nearest integer. Each audio frame is therefore represented by one pattern of length $T$ time-steps containing 128 impulses (spikes), one for each feature coefficient $k$. 
This encoding exploits the fact that the network dynamics (eqs.\ref{eqn:discreteLIF}) are invariant to a global shift of the input pattern in time, therefore to a global shift in coefficients magnitude, implying that the information is conveyed in the synchronicity between spikes of different input neurons rather than the absolute timing of each spike. This allows for a minimal description of the input vector with 128 events or spikes per audio frame, achieving high sparsity.

\vspace{-2pt}
\subsection{Experiments}

We present two different architectures that can tackle the problem, both take as input a feature vector corresponding to a $64$ms audio frame encoded into a spiking pattern of $128$ input neurons, $T=100$ timesteps and $\Delta T = 1$. Spikes are sent to the hidden layers in a fully-connected feedforward fashion and reach the output layer composed of two neurons: one coding for voice and the other for no voice.

Neurons in the network follow eqs.\ref{eqn:discreteLIF} with the exception that output neurons do not have an activation function, namely, they do not spike and do not get reset. We define a loss function on the maximum voltage reached by the output neurons, specifically on the cross-entropy between the softmax of the two maximum values and the true labels ($[1,0]$ for 'no-speech' and $[0,1]$ for 'speech'). For both the experiments we minimized the loss function using the Adam optimization method \cite{Kingma2014AdamAM} (learning rate = $10^{-4}$, batch size = 256). The classified label is given by the difference between the maxima of the voltage traces of the output neurons: if $\max(V_{speech}) - \max(V_{no\_speech}) > \rho$, the frame is labelled as speech and vice-versa for no-speech. $\rho$ can be set to zero or tuned to bias the classifier towards one class.

The first network architecture, SNN h1, is composed of one hidden layer of $200$ neurons with $\tau_{syn}=5$, $\tau_{mem}=10$. It classifies each audio frame independently from the others. To overcome the lack of contextual information, the predicted labels (0 for no-speech and 1 for speech) are smoothed with a median filter of size 11, consisting of 224ms of audio.

The second architecture, SNN h2, has been designed to include past contextual information about the classified audio frame. The network is fed with 5 successive frames, which amount to 500 time-steps, and asked to classify the last. To include a longer memory effect in the neurons a second hidden layer of 15 neurons with $\tau_{mem}=300$ is introduced after a hidden layer of 100 neurons and $\tau_{mem}=10$ while $\tau_{syn}=5$ for all neurons, consisting of 14.330 weights. The classification is performed on the maximum values of the output neurons voltage while the last frame is presented. The same median filter is then applied to the predicted labels. 

An implementation of the networks can be found in \cite{code}.

\subsection{Pruning}
\label{sec:prun}
Deployment of the described architectures on neuromorphic chips need to account for architectural constraints of the chip's internal routings. One major constraint is the maximum number of input/output connections that each neuron can have in the chip. Moreover, reducing the number of spikes received by each neuron, Synaptic Operations (SOPs), reduces the power consumption of the chip. 
Pruning techniques for spiking networks have been proposed in \cite{bellec2018deep}, we tested instead whether the Lottery Ticket Hypothesis \cite{frankle2018the} developed for ANNs applies also to our SNN h1 spiking model. The proposed technique consists in training the network multiple times starting from the same initialization weight vector: at the end of each iteration (which consists of a full training procedure) the connections which weights are closest to zero gets pruned, the successive training iteration is initialized with the same initialization vector but pruned of the connections found in the preceding iteration. This method allows for direct control on the number of connections pruned and shows limited to no accuracy loss in many ANN models \cite{frankle2018the}.

\section{results}
\label{sec:majhead}
The described models (SNN h1 and SNN h2) are trained and tested on the QUT-NOISE-TIMIT \cite{dean2010qut} corpus, developed for testing VAD applications in noisy scenarios. It contains 600 hours of audio tracks with different levels and types of background noise mixed with the TIMIT \cite{zue1990speech} clean speech corpus. The dataset is divided into two parts, group A and B, in which same noise types have been recorded in different locations. We trained the models on the former and tested them on the latter, each of them totalled around $16.87 \times 10^6$ audio frames.

\begin{table}[t]
\centering
\begin{tabular}{@{}lcccccc@{}}
\toprule
\textbf{Method / SNR} & \textbf{+15} & \textbf{+10} & \textbf{+5} & \textbf{0} & \textbf{-5} & \textbf{-10} \\ \midrule
\textbf{Sohn\cite{sohn1999statistical}} & 11.1 & 13.4 & 19.7 & 25.9 & 31.3 & 37.6 \\
\textbf{Segbroeck\cite{van2013robust}} & 6.1 & 6.0 & 10.4 & 10.8 & 18.3 & 23.2 \\
\textbf{Neurogram \cite{jassim2018voice}} & 5.5 & 5.9 & 10.2 & 10.0 & 17.5 & 23.7 \\ \midrule
\textbf{SNN h1\textit{-w}} & 2.9 & 4.5 & 7.1 & 9.7 & 12.5 & 16.3 \\
\textbf{SNN h2\textit{-w}} & 5.0 & 5.8 & 7.3 & 9.6 & \cellcolor[HTML]{C0C0C0}12.2 & \cellcolor[HTML]{C0C0C0}15.7 \\
\textbf{SNN h1} & \cellcolor[HTML]{C0C0C0}2.4 & \cellcolor[HTML]{C0C0C0}3.4 & \cellcolor[HTML]{C0C0C0}5.9 & 10.2 & 16.3 & 26.5 \\
\textbf{SNN h2} & 3.9 & 4.5 & 6.2 & \cellcolor[HTML]{C0C0C0}9.4 & 14.1 & 21.1 \\ \bottomrule
\end{tabular}
\caption{DCF\% scores comparison adapted from \cite{jassim2018voice}. In grey the best performing model for each SNR (dB). Models with suffix $-w$ are trained with a weighted loss function to match the DCF metric, lack of suffix refers to a balanced loss.}
\vspace{-6pt}
\label{table:dcf}
\end{table}

Following the protocol suggested by the dataset \mbox{creators} \cite{dean2010qut}, we grouped the SNR levels into low (15dB, 10dB), medium (5dB, 0dB) and high (-5dB, -10dB) and trained a different model for each noise level group. Training a single model on all SNR levels outperforms the models trained solely for medium and high noise, suggesting that knowledge of all levels is beneficial to the final performance and being SNR agnostic is most suitable for real-world applications. 

We compared to the following standard untrained techniques: advanced front-end \cite{li2004complexity} (ETSI), ITU-T G.729 Annex B \cite{benyassine1997itu} (G729B), Likelihood Ratio test \cite{sohn1999statistical} (Sohn) and Long Term Spectral Divergence \cite{ramirez2004efficient} (LTSD). Recent trained methods, producing different models for each noise levels, are also compared: GMM trained on Mel Frequency Cepstral Coefficients \cite{ng2012developing} (GMM-MFCC), Complete Linkage Clustering \cite{ghaemmaghami2015complete} (CLC), ANNs (Segbroeck, Neurogram) \cite{van2013robust, jassim2018voice} and a Convolutional Neural Network \cite{silva2017exploring} (CNN).

Table \ref{table:dcf} compares models trained on all noise levels and evaluated on the Detection Cost Function, $DCF = 0.25 FAR + 0.75 MR$, MR and FAR being miss and false alarm rates, extending the table of results from Neurogram \cite{jassim2018voice} and showing substantial improvements.

\begin{figure}[t]
\centering
\includegraphics[width=8.5cm]{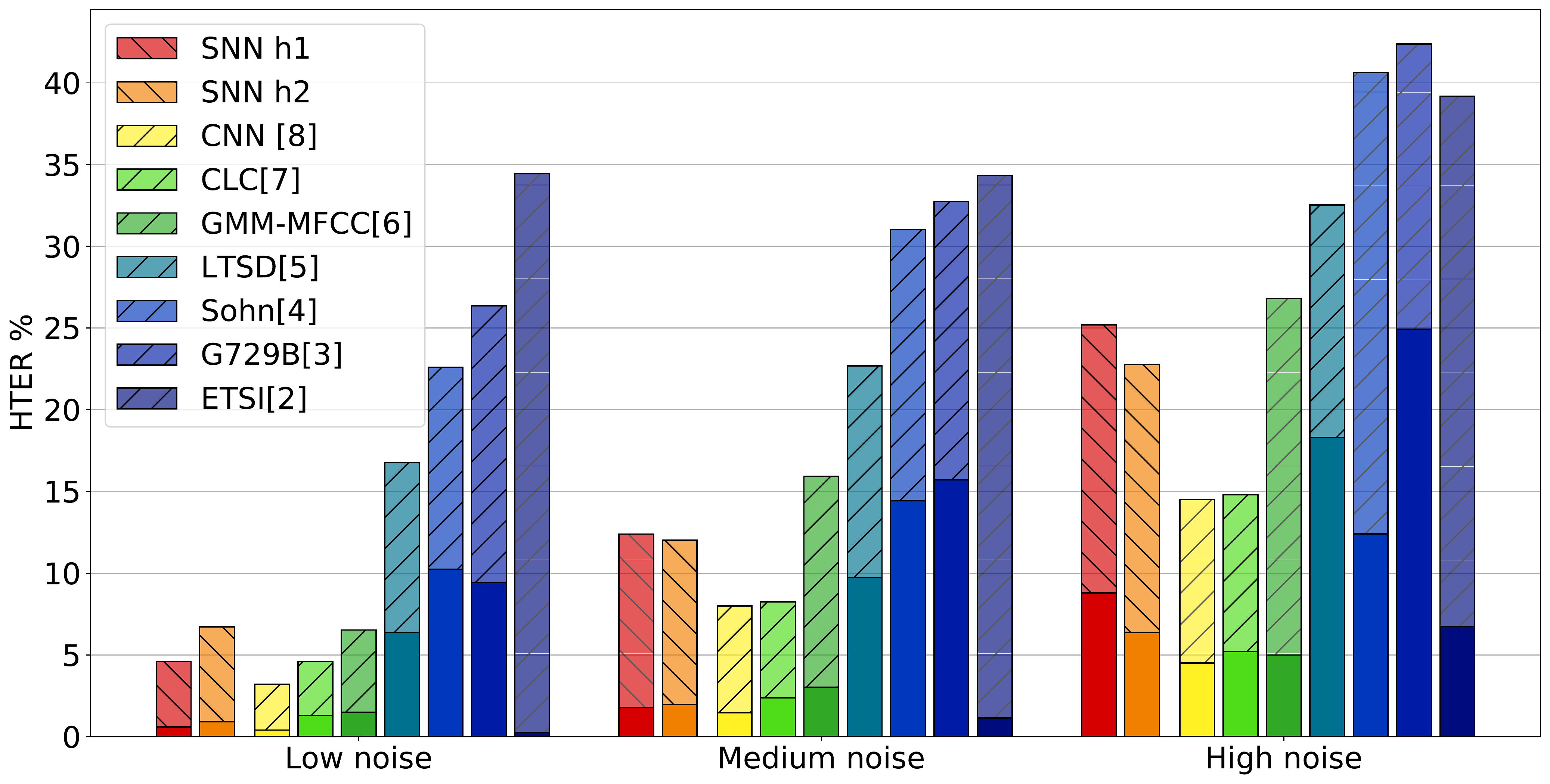}
\caption{Comparisons of HTER\% values for different noise levels. Dark and light patterned shades highlight $MISS$ and $FA$ contributions, respectively. Legend order, top to bottom, is respected in the bars' order, left to right. }
\vspace{-6pt}
\label{hter}
\end{figure}

\begin{figure}[t]
\centering
\includegraphics[width=8.5cm]{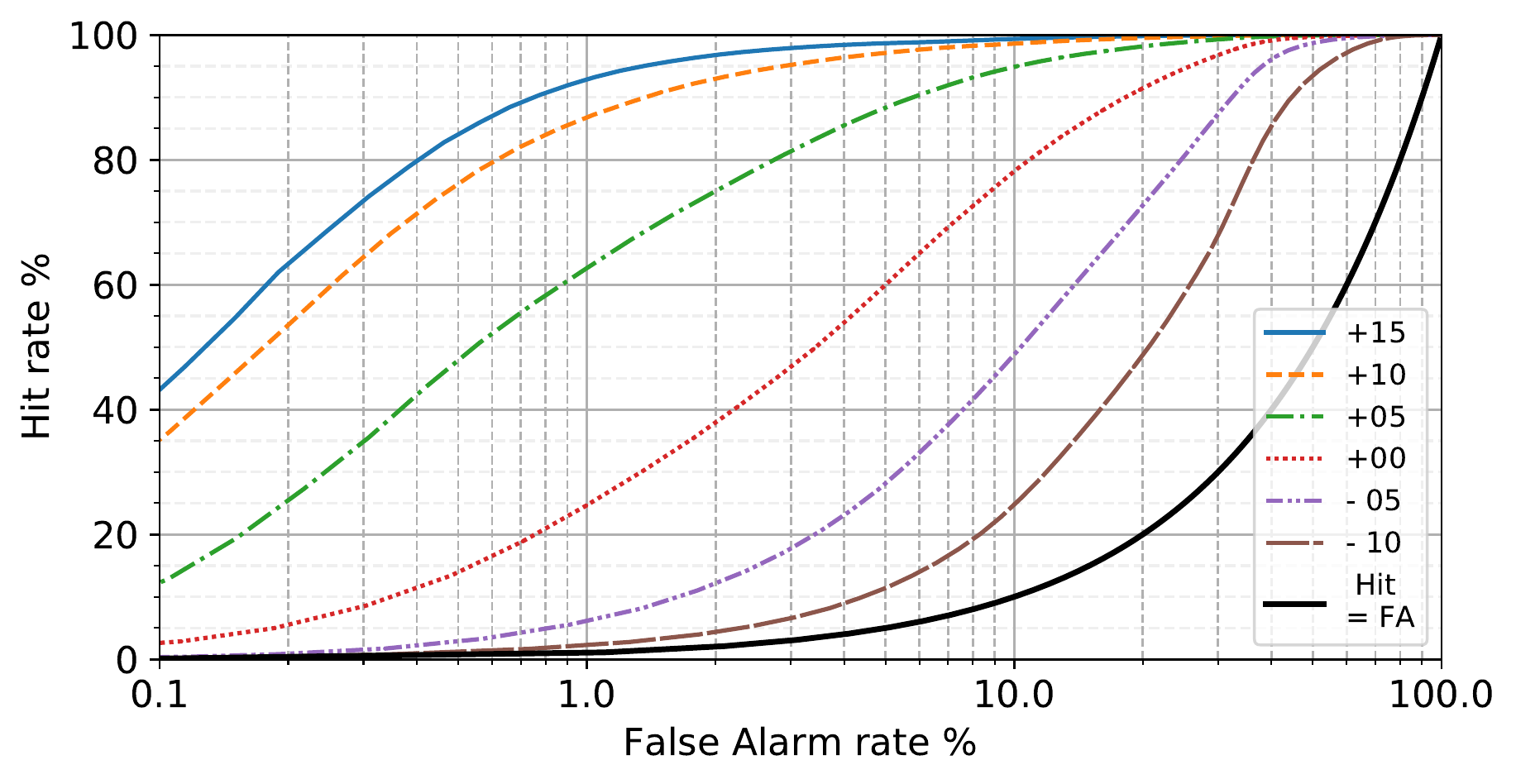}
\vspace{-5pt}
\caption{ROC curve for the model SNN h1, the lower black thick line corresponds to Hit rate = False Alarm rate.}
\vspace{-6pt}
\label{rocshort}
\end{figure}

The results shown in fig.\ref{hter} compare our SNNs with the protocol previously mentioned evaluated with the Half Total Error Rate: $HTER = 0.5 MR + 0.5 FAR$. The proposed one-hidden layer model, SNN h1, scores 4.6\%, 12.4\% and 25.2\% for low, medium and high noise levels, respectively, being able to compete with trained models and outperforming the standard untrained models. The second model proposed, with two hidden layers and different membrane time constants, SNN h2, scores 6.7\%, 12.0\% and 22.7\%, achieving slightly better performances in medium and high noise scenarios but losing performance on low noise levels.
The models' classification outcomes can be biased with the parameter $\rho$ to tailor for a specific application and compute the Receiver Operating Characteristic (ROC) curve in fig. \ref{rocshort}.

\begin{table}[b]
\vspace{-5pt}
\centering
\begin{tabular}{@{}l@{\hskip 7pt}cccc@{}}
\toprule
\multicolumn{1}{c}{} & Rate & \#Params      & Power & HTERs \%              \\  \midrule 
SNN h1               & $1$ - $4.7$       & $26k$ & $33.0 \, \mu W$ & $4.6 \| 12.4 \| 25.2$ \\
SNN h1-$p$             & $1$ - $2.9$       & $4096$ & $25.1\, \mu W$ & $4.7 \| 12.5 \| 25.8$ \\ \bottomrule
\end{tabular}
\caption{Comparison between SNN h1 and its pruned version. Average neuron spiking rate per inference is shown for the input and hidden layers, the number of parameters and estimated power consumption follow. The three HTER values correspond to low, medium and high noise respectively.}
\label{table:energy}
\end{table}

The pruning procedure described in section \ref{sec:prun} consists of 4 iterations in which the connections from input to hidden layer of the SNN-h1 model are pruned up to 15\% (with steps of 70\%, 40\%, 20\%, 15\%). The pruned model, SNN h1$-p$, does not lose any significant performance with respect to the original fully connected model, suggesting that the Lottery Ticket Hypothesis holds even for these types of networks. 
Power consumption is strongly dependent on the type of chip and network chosen. Lacking such hardware, our estimates follow the benchmark results of TrueNorth \cite{merolla2014million} and must be interpreted as indicative values. Following fig.4B in \cite{merolla2014million} we find the total TrueNorth power consumption of 105 (80) $mW$ for SNN h1(SNN h1-$p$) at 3.7 (1.8) average spiking rate and 80 (13) average active synapses per neuron. Dividing by the total number of neurons in the TrueNorth chip ($4096 \times 256$) and multiplying by the ones in our networks we estimate the power consumption of table \ref{table:energy}.
The pruned network achieves the same performance as SNN h1 consuming $25$\% less energy with 40\% less spiking activity. The power estimates in table \ref{table:energy} do not include feature preprocessing and encoding and assume that total power consumption in the TrueNorth chip scales linearly with the number of neurons. Dedicated chips, with advanced mixed analogue-digital implementations, might lower even more these estimates.
Current low power VAD implementations reach less than $1 \mu W$, but at a performance cost: e.g. in \cite{cho201917} babble noise at 5dB scores 84\% and 72\% on hit rate and correct rejections whereas similar conditions ("CAFE" of \cite{dean2010qut}) on SNN h1 reach 97\%, 84\%; other power comparisons can be found in table 1 of \cite{cho201917}.

\section{Conclusions and limitations}
\label{sec:page}
We proposed a novel VAD application using an SNN model and we showed that it competes with state of the art algorithms while keeping a very low energy consumption. We addressed the connectivity limitations of neuromorphic chips by testing the lottery ticket hypothesis \cite{frankle2018the} with positive results, achieving even less energy consumption. 
Previous studies \cite{bellec2018long, zenkePoster, mostafa2017supervised} applied similar models to simple image datasets, where the features precise values are not necessary to solve the task, namely the task can be solved by using only binary features \cite{mostafa2017supervised}. The novelty of the proposed work lies in showing how these models are able to process more complex continuous features such as Mel filterbank coefficients in a real-world application, highlighting the power of the method and the high level of sparsity these features can be represented with temporal coding, in contrast with ANN to SNN conversion methods that rely on spike rates to convey information \cite{rueckauer2017conversion} such as the spiking VAD solution proposed in \cite{essera2016convolutional} at 26.1 mW. 
The main limitation of the model is the computationally heavy training procedure that constrains hyperparameter optimization. We reached competitive performance by using basic features and encoding schemes, room for improvement can be found by exploring more elaborate architectures, features, and their conversion into spike patterns.
With this example we want to stimulate further research on these models, applied to different tasks, suggesting that they can compete with embedded deep learning approaches with a good trade-off between energy consumption and performance.

\bibliographystyle{IEEEbib}
{\ninept \bibliography{strings,refs}}

\end{document}